\documentclass[letterpaper, 10 pt, conference]{ieeeconf}  

\IEEEoverridecommandlockouts                              

\usepackage{graphics} 
\usepackage{epsfig} 
\usepackage{mathptmx} 
\usepackage{times} 
\usepackage{amsmath} 
\usepackage{amssymb}  
\newtheorem{definition}{Definition}
\usepackage[ruled,linesnumbered]{algorithm2e} 
\usepackage{amsmath}
\usepackage{graphicx}
\usepackage{booktabs} 
\usepackage{color,soul}
\usepackage{diagbox}
\usepackage{makecell}
\usepackage{multirow,multicol}
\usepackage{textcomp}
\usepackage{multicol}
\usepackage{multirow}
\usepackage{tabularx}
\usepackage{makecell}
\usepackage{colortbl}
\usepackage{caption}
\usepackage{subfigure}
\usepackage{lipsum}
\usepackage[utf8]{inputenc}
\usepackage{algorithm2e}
\usepackage{array}
\newcommand{\PreserveBackslash}[1]{\let\temp=\\#1\let\\=\temp}
\newcolumntype{C}[1]{>{\PreserveBackslash\centering}p{#1}}
\newcommand{\ours}{\texttt{PSAD}\xspace}

\title{\LARGE \bf Personalized State Anxiety Detection: An Empirical Study with Linguistic Biomarkers and A Machine Learning Pipeline}

\author{Zhiyuan Wang$^{1}$, Mingyue Tang$^{1}$, Maria A. Larrazabal$^{2}$, Emma R. Toner$^{2}$, Mark Rucker$^{1}$, Congyu Wu$^{3}$, \\Bethany A. Teachman$^{2}$, Mehdi Boukhechba$^{1}$ and Laura E. Barnes$^{1}$
\thanks{*This work was partially supported by a 3Cavaliers Seed Grant and by the National Institute of Mental Health of the National Institutes of Health under award number R01MH132138.}
\thanks{$^{1}$Zhiyuan Wang, Mingyue Tang, Mark Rucker, Mehdi Boukhechba, and Laura Barnes are with the Department of Systems and Information Engineering,
        University of Virginia, Charlottesville, VA, USA
        {\tt\small \{vmf9pr,utd8hj,mr2an,mob3f,lb3dp\}@virginia.edu}}%
\thanks{$^{2}$Maria A. Larrazabal, Emma R. Toner, and Bethany Teachman are with the Department of Psychology, University of Virginia, Charlottesville, VA, USA
        {\tt\small \{ml4qf,ert6g,bat5x\}@virginia.edu}}%
\thanks{$^{3}$Congyu Wu is with the Department of Systems Science and Industrial Engineering, College of Engineering and Applied Science, Binghamton University, Vestal, NY, USA
        {\tt\small congyu.wu@binghamton.edu}}%
}

\begin{document}

\maketitle
\thispagestyle{empty}
\pagestyle{empty}

\begin{abstract}
Individuals high in social anxiety symptoms often exhibit elevated state anxiety in social situations. Research has shown it is possible to detect state anxiety by leveraging digital biomarkers and machine learning techniques. However, most existing work trains models on an entire group of participants, failing to capture \emph{individual differences} in their psychological and behavioral responses to \emph{social contexts}. To address this concern, in Study 1, we collected linguistic data from N=35 high socially anxious participants in a variety of social contexts, finding that digital linguistic biomarkers significantly differ between evaluative vs. non-evaluative social contexts and between individuals having different trait psychological symptoms, suggesting the likely importance of personalized approaches to detect state anxiety. In Study 2, we used the same data and results from Study 1 to model a multilayer personalized machine learning pipeline to detect state anxiety that considers contextual and individual differences. This personalized model outperformed the baseline’s F1-score by 28.0\%. Results suggest that state anxiety can be more accurately detected with personalized machine learning approaches, and that linguistic biomarkers hold promise for identifying periods of state anxiety in an unobtrusive way.
\end{abstract}


\section{Introduction}

Social anxiety disorder (SAD) is highly prevalent, impacting 13\% of adults in the United States at some point in their lifetime \cite{kessler2012twelve}. Individuals with SAD fear and often avoid social interactions, or endure them with significant anxiety \cite{hofmann2007cognitive}. However, roughly 80\% of individuals delay or avoid treatment \cite{grant2005epidemiology,wang2005failure}. Digital interventions delivered to individuals via mobile technology (e.g., smartphones) in daily life can increase treatment access among individuals with SAD. “Just-in-time” adaptive interventions (JITAIs; \cite{nahum2018just}) are a good candidate to increase access to care. By providing individuals with treatment components when and where they need them, JITAIs have the potential to help socially anxious individuals navigate social situations more effectively in real time. Crucially, to deploy JITAIs, we need to detect when \emph{socially anxious} people are in need of an intervention.

Machine learning (ML) offers a computational solution to detect state anxiety status from digital biomarkers (i.e., passively sensed bio-behavioral indicators) \cite{orru2020machine}. In particular, using ML with linguistic biomarkers is a promising option for understanding social anxiety as this data can be passively collected without active user input and contains a wealth of psychological information \cite{hofmann2012linguistic}.

Despite this potential, a limitation of much ML-based computational work with psychological data is its reliance on nomothetic approaches. Specifically, it has generally used one-size-fits-all approaches (e.g., modeling and reasoning about entire observed populations indiscriminately) \cite{dadi2021population}, even though idiographic (i.e., person-specific) outcomes are typically of most interest to clinicians (e.g., a specific individual's risk of developing an anxiety disorder) \cite{wang2022personalized}. At the same time, most personalized ML approaches require extensive data points from individuals, which is often impractical to collect (e.g., due to participant burden, low engagement, time and resource limitations) \cite{olfson2000barriers}. There are thus many challenges for ML models to efficiently and effectively detect state anxiety (e.g., model overfitting, inaccuracy, and bias) \cite{vabalas2019machine}.

This study aims to develop a \emph{personalized ML pipeline} that accounts for contextual and individual information to detect state anxiety. To this end, we propose a two-stage study. Study 1, an \emph{empirical study}, examines the differences in linguistic biomarkers associated with state anxiety across social contexts and individual subgroups. In controlled dyadic Zoom conversations, we observe that socially anxious college students' (N=35 final sample) linguistic patterns significantly differ across experimentally manipulated contexts (one designed to be explicitly socially evaluative and one not explicitly socially evaluative) and clustered psychological-symptom severity subgroups. Based on these findings, in Study 2, we develop and test a \emph{personalized ML pipeline} to detect state anxiety that accounts for contextual and individual differences using multi-layer fine-tuning training approaches. This pipeline hierarchically trains the model at the population, contextual, and individual levels by progressively adding new neural network layers and narrowing down the data samples grouped by contexts and individuals.

In summary, this paper makes the following contributions:
\begin{itemize}
    \item We report evidence of significant linguistic differences between situational contexts and individual subgroups among socially anxious individuals, which suggests idiographic distinctions and motivated our ML models.
    \item We propose a multilayer personalized ML pipeline to detect state anxiety, which is able to hierarchically fine-tune a population-based model to capture contextual (i.e., social threat) and subgroup-based (i.e., psychological symptom severity) domain knowledge.
\end{itemize}

\section{Empirical Study} \label{section4-case-study}

\subsection{Study Design}
All study procedures were approved by the Institutional Review Board (IRB) of a large U.S. university and conducted under the supervision of a licensed clinical psychologist and researcher with expertise in anxiety disorders. We recruited 45 undergraduate participants with a Social Interaction Anxiety Scale (SIAS) score of 34 or above, indicating greater trait social anxiety (scale ranges from 0 to 80). Ten of the 45 participants did not complete the social experiences to be analyzed in this paper, leaving 35 participants eligible. The 35 participants had a mean age of 19.46 (SD = 2.09), and the majority were female (74.3\%) and White (82.8\%). 

In the broader parent study, participants completed a series of social and non-social tasks involving different group sizes (i.e., alone, pairs, and groups of 4-6) and levels of experimenter-manipulated social threat. All study procedures were conducted virtually via Zoom. For the purposes of the present investigation, we focus on the two conversation tasks that participants completed in pairs (because the dyadic exchange allowed for clear evaluation of linguistic features). One conversation was explicitly evaluative and the other was \emph{not} explicitly evaluative based on experimenter instructions (termed evaluative and non-evaluative, accordingly, for ease of reference). Specifically, we manipulated the level of social-evaluative threat present in the conversations: Prior to one conversation, participants were told that their partner would rate their social performance following the conversation (i.e., \emph{evaluative}); prior to the other conversation, participants were told that they would \textit{not} be rated (i.e., \emph{non-evaluative}). During the two conversations, two participants discussed a randomly assigned topic for four minutes. The order of the two conversations was randomized.

\subsection{Data Collection}
Participants reported on their baseline and concurrent state anxiety (Def \ref{define:baseline-concurrent}) during each of the social experiences via two brief surveys to report state anxiety levels (Def \ref{define:state-anxiety}). 
\begin{definition} \label{define:baseline-concurrent}
  {\bf Baseline and Concurrent State Anxiety}: Baseline state anxiety refers to the state anxiety level that participants report prior to learning about the upcoming social experience they will complete. Concurrent state anxiety refers to participants' most intense state anxiety \emph{during} the social experience, as reported by each participant immediately after the experience ended. In this study we assessed participants' state anxiety via a Qualtrics survey by asking them to rate how anxious they felt on a five-point Likert scale from Very Calm (1) to Very Anxious (5).
\end{definition}

\begin{figure}[]
\centering
\includegraphics[width=0.75\columnwidth]{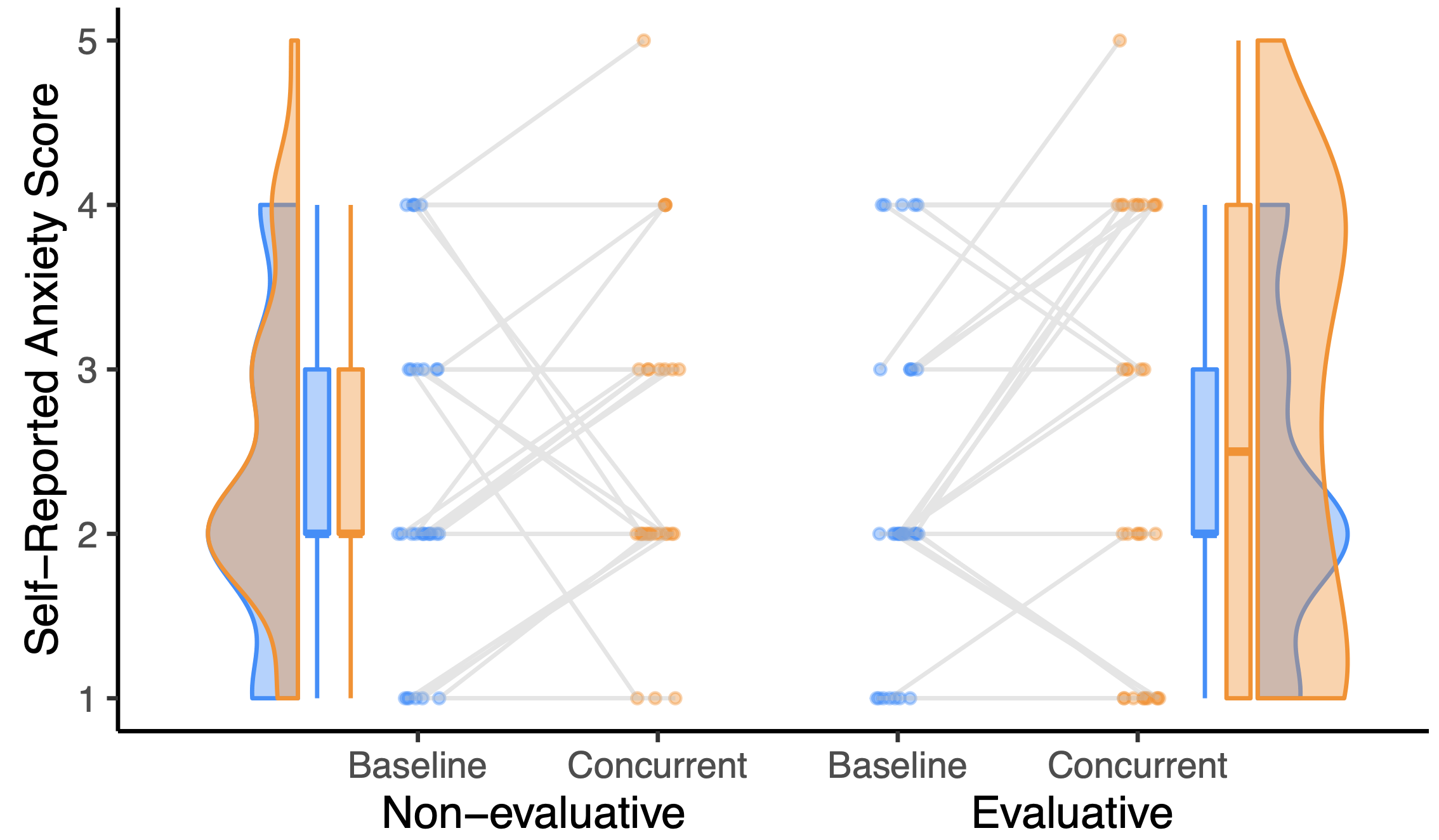}
\caption{Self-reported baseline v.s. concurrent state anxiety. Each point reflects an anxiety score reported by a participant.}
\label{fig:groundtruth}
\end{figure}

\begin{definition} \label{define:state-anxiety}
  {\bf State Anxiety Status}: In this study, we aim to detect \emph{state anxiety} status (i.e., status when the participants were in \emph{high and/or elevated} state anxiety), as both represent times when individuals may benefit from a JITAI. Specifically, \emph{high} state anxiety refers to periods when a participant reports either feeling anxious (4) or very anxious (5) during the concurrent stage. \emph{Elevated} state anxiety refers to periods when the participants' concurrent anxiety is higher than their baseline anxiety, regardless of the particular score reported.
\end{definition}
See Figure \ref{fig:groundtruth}, state anxiety oscillated during the study with notable elevation in the evaluative contexts (the right half).

Participants' \emph{psychological symptoms} (in this case, self-reported trait anxiety and depression symptoms, and emotion regulation) were measured at the end of the study by four scales. The scales included the depression subscale from the Depression, Anxiety and Stress Scale (DASS; \cite{lovibond1996manual}) \cite{ruscio2008social}, the Social Interaction Anxiety Scale (SIAS; \cite{mattick1998social}), Brief Fear of Negative Evaluation scale (BFNE; \cite{leary1983brief}), and Difficulties in Emotion Regulation Scale - Short Form (DERS-SF; \cite{kaufman2016difficulties}).

The audio of the participants was recorded through Zoom. Otter.ai then transcribed the conversation text, identifying the start and end time points and the speaker for each sentence. The identified time points were used to segment and aggregate each participant's recordings sentence by sentence. Altogether, 55 samples of audio recordings and corresponding self-reported state anxiety were collected; 20 participants completed both the non-evaluative and evaluative sections (40 samples); due to the time limitation of the data collection, 7 participants only completed the non-evaluative experience (7 samples), and 8 participants only completed the evaluative part (8 samples).

\subsection{Linguistic Biomarkers}

\begin{table}[htbp]
\centering
\caption{Linguistic feature list with descriptions.}
\resizebox{0.48\textwidth}{!}
{
\small
\begin{tabular}{l p{2.2cm} p{6cm}}
\toprule
Domain & Feature & Description \\ \midrule
\multirow{4}{*}{Acoustic}
& Pitch & Mean and delta of voice frequency \\
& Energy & Mean and delta of voice intensity \\
& Zero-crossing rate & Mean and delta of the number of times the voice signal crosses the zero-axis \\
& Spectral center & Mean and delta of the voice spectrum center \\ \midrule

\multirow{3}{*}{Syntactic} & Avg. word count & Average number of words per sentence \\
& Long sentence  & Percentage of sentences with $>$ 15 words \\
& Sentence count & Total number of sentences spoken \\ \midrule

\multirow{6}{*}{Lexical} & Pos. emotion & Number of words indicating positive sentiment \\
& Neg. emotion & Number of words indicating negative sentiment \\
& \emph{I}-statements & Number of first-person pronouns \\
& \emph{You}-statements & Number of second-person pronouns \\
& Negations & Number of negation words or phrases \\
& Stop words & Number of function words \\ \bottomrule
\end{tabular}%
}
\label{table:linguistics}
\end{table}

We extracted linguistic biomarkers from speech audio and transcripts in each sample, as a large body of literature has linked anxiety with linguistic biomarkers \cite{hofmann2012linguistic,sonnenschein2018linguistic}. 
We extracted \emph{phonetic}, \emph{syntactic}, and \emph{lexical} domain features for each sample (Table \ref{table:linguistics}). To effectively model and measure participants' linguistic behaviors, we only extracted one-dimensional features verified by the existing literature.

\subsection{Results}

\subsubsection{\textbf{Situational Context Matters}} \label{sec:result-context}

First, we explored the differences in linguistic biomarkers between \emph{non-evaluative v.s. evaluative} situational contexts with paired analysis on the 20 participants who completed both non-evaluative and evaluative experiences. 

\begin{figure}[!t]
\centering
\includegraphics[width=0.9\columnwidth]{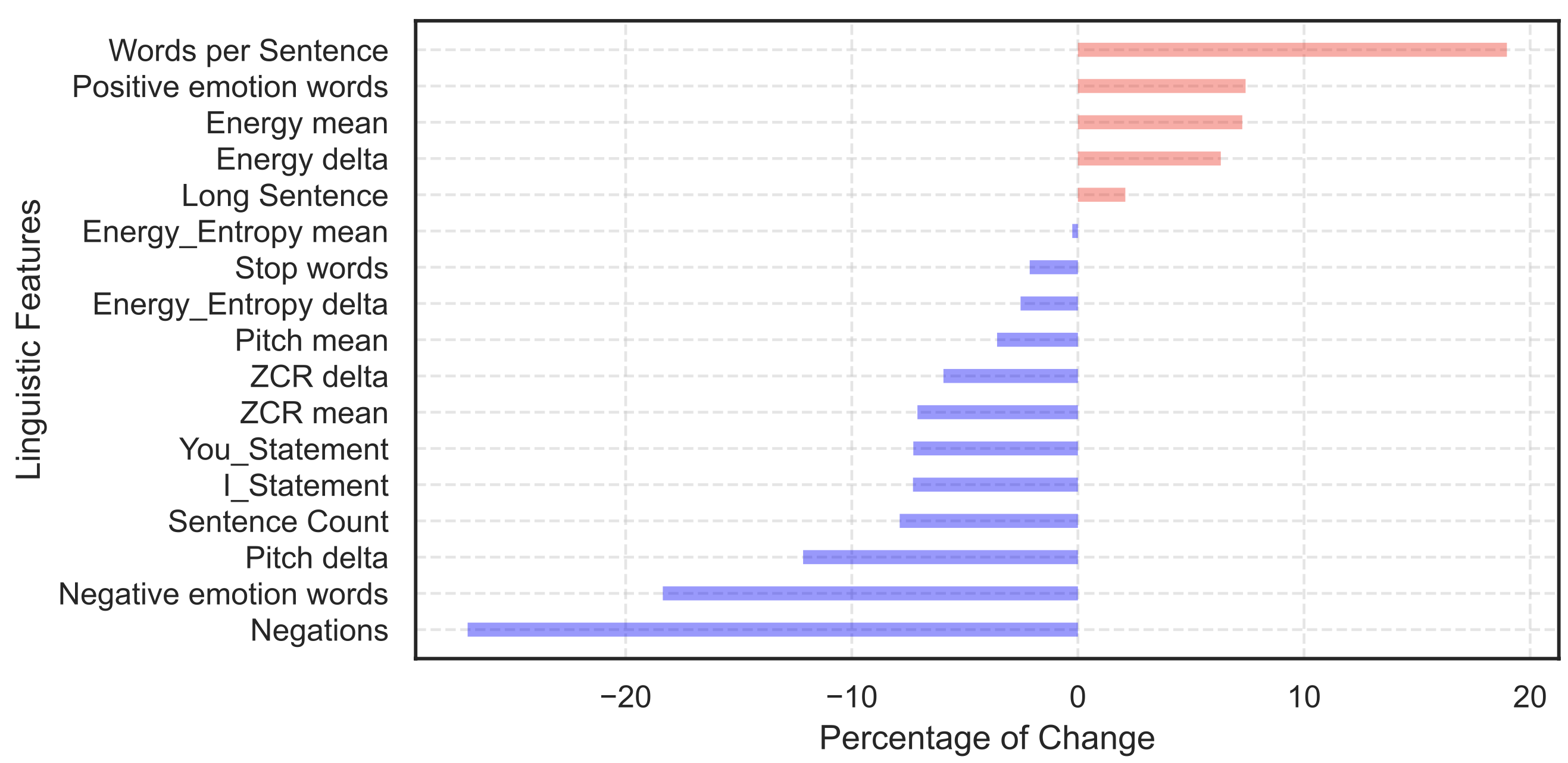}
\caption{Percentage loss (in blue) vs. gain (in red) for linguistic biomarkers between non-evaluative vs evaluative context.}
\label{fig:non_eval}
\end{figure}

After normalizing each feature to a scale of 0-1 for comparison on a common scale, using paired analysis, we compared the percentage changes of linguistic biomarkers in the evaluative contexts compared to the non-evaluative contexts (see Figure \ref{fig:non_eval}). In the evaluative context, notably decreased linguistic biomarker scores included negations (-27.0\%, p=0.306), pitch delta (-22.5\%, p=0.018), negative emotion words (-18.4\%, p=0.066), I-statements (-14.6\%, p=0.396), zero-crossing rate (-11.7\%, p=0.100), sentence count (-11.2\%, p=0.087), zero-crossing rate delta (-9.6\%, 0.050), and pitch mean (-8.7\%, p=0.096). Increased biomarker scores between evaluative and non-evaluative contexts included words per sentence (+34.5\%, p=0.015) and energy delta (+9.5\%, p=0.171). The Wilcoxon signed rank test was used to test the p-values with p$<$.05 considered a reliable effect.

\subsubsection{\textbf{Individual Differences Matter}} \label{sec:result-individual}

 We then performed cluster analysis to aggregate participants according to their trait psychological scales by clustering the participants by their DASS, SIAS, BFNE, and DERS scales using the K-means algorithm. Individuals with similar trait symptom severity were gathered (i.e., similar score patterns on social anxiety and depression severity, and emotion regulation tendencies; henceforth called \emph{symptom severity} for ease of reference \footnote{Though we recognize DERS reflects a transdiagnostic vulnerability for emotional disorders, rather than a measure of symptom severity per se.}). To find optimal parameter $K$ of K-means, we calculated the silhouette scores of K values from 2 to 5. The optimal $K$ found is 2 (silhouette score=0.394), while the silhouette scores are 0.343, 0.373, and 0.364 when $K$ are 3, 4, and 5. Selecting $K=2$ to operate cluster analysis, as profiled in Table \ref{table:psychological-measures-profile}, the two clustered individual cohorts reflected two symptom severity groups. Group 1 (named high symptom severity) generally had high symptoms of social anxiety and depression (DASS and SIAS), fear of negative evaluation (BFNE) and difficulty with emotion regulation (DERS), whereas group 2 (named low symptom severity) had lower scores on those measures.

\begin{table}[!t]
\centering
	\footnotesize
	\renewcommand{\arraystretch}{1}
	\centering
	\caption{Profile of the 4 psychological scales among the two symptom groups. Abbreviation: Symptom=sx.}
	\label{table:psychological-measures-profile}
    \begin{tabular}{ l c c c c}
		\toprule
		\bfseries{Measures}	& \multicolumn{2}{c}{\bfseries{ \ \ High Sx (N=13)  \ \ }} 	& \multicolumn{2}{c}{\bfseries{\ \ Low Sx (N=17) \ \ }}   	\\
							\cmidrule(lr){2-3} 		\cmidrule(lr){4-5} 	 
										
										    &  mean 	&  std  & mean  &  std 		\\

		\midrule
					\textbf{DASS}              &  69.16 & 10.31   & 51.13 & 7.67      \\ 
                    \textbf{SIAS}               &  33.44 & 4.51   & 23.93 & 8.00            \\
                    \textbf{BFNE}               &  58.12 & 6.27   & 41.93 & 5.75          \\
                    \textbf{DERS}               &  15.36 & 5.05   & 12.27 & 5.02         \\
                    
		\bottomrule
	\end{tabular}
\end{table}

\begin{figure}
	\centering
	\subfigure[Non-evaluative context]{
		\begin{minipage}{0.5\textwidth}
             \includegraphics[width=\textwidth]{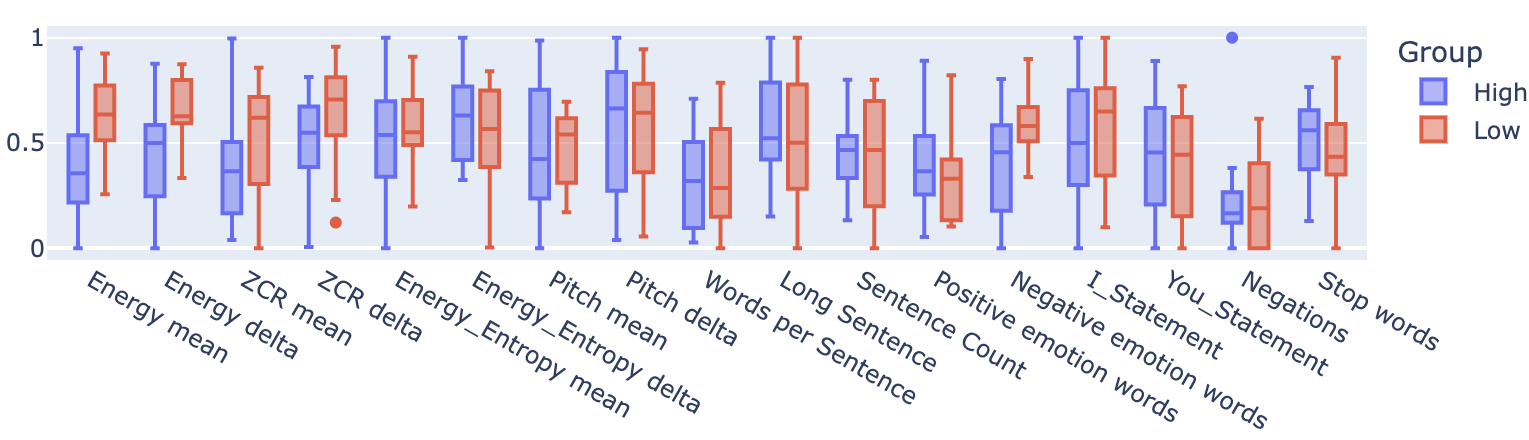} \\
		\end{minipage}
	}

	\subfigure[Evaluative context]{
		\begin{minipage}{0.5\textwidth}
			\includegraphics[width=\textwidth]{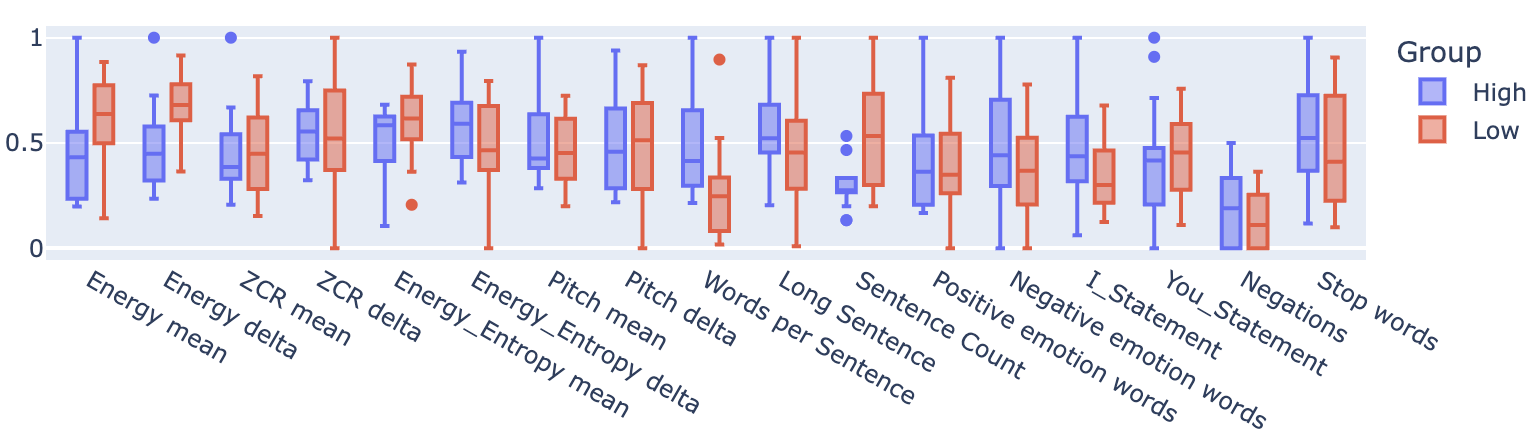} \\
		\end{minipage}
	}
	\caption{Distributions of 0-1 normalized linguistic features of clustered symptom severity subgroups in non-evaluative and evaluative contexts, respectively.} 
	\label{fig:linguistic-clustered-group}
\end{figure}

As shown in Figure \ref{fig:linguistic-clustered-group}, the two subgroups behaved differently in some aspects in linguistic. Specifically, the high symptom group generally had \emph{lower} energy mean and delta, zero-crossing rate mean, energy entropy mean, pitch mean, as well as \emph{higher} energy entropy delta, words per sentence, long sentence rate, and stop word rate in both contexts. Typically, in the evaluative context, the high symptom group had more \emph{words per sentence}. Also, in the evaluative context, individuals with high symptom levels tend to use more negative emotional words, perhaps indicating a higher degree of distress, compared to those with low symptom levels. Moreover, high symptom individuals used more first-person pronouns than the low symptom group in the evaluative context. These patterns suggest that, in addition to individual and subgroup differences in behavior across contexts, certain behavioral patterns are more likely to be associated with evaluative social contexts than others.

\section{Personalized Learning Pipeline} \label{section5-model}

Sections \ref{sec:result-context} and \ref{sec:result-individual} indicated there are behavioral differences between the two contexts and the two cohorts, suggesting that a personalized ML pipeline should ideally be tailored to the variability in social contexts and across persons/cohorts. We devised such a ML pipeline based on Multilayer Perceptron (MLP) model \cite{ramchoun2016multilayer}, which we term the Personalized State Anxiety Detector (\textbf{\ours}).

\subsection{Problem Formulation}

Given the raw input data $D_p = \{d_{1}, d_{2}, \cdots, d_{n}\}$, where $n$ indicates the number of linguistic views of $D_p$, of participant $p$, the task is to predict binary state anxiety status $S_p$ (Definition \ref{define:state-anxiety}). To achieve this, we proposed a framework that includes: 1) a set of biomarker extractors $\{f_{1}, f_{2}, \cdots, f_{n}\}$ to map different views of features into spaces $F^{p}_i \in \mathbb{R}^{m_i}$, where $m_i$ is the dimension of features in view $i$; 2) a multiview fusion method that fused $F^p_1, F^p_2, \cdots, F^p_n$ into a hidden vector $H_p$; 3) applies a personalized classification model $C_{p}$ to $H_p$ that can be narrowed down by participant $p$, leveraging the contextual and individual information $SC_p$ to predict $S_p$.

\subsection{Personalized ML Pipeline Design}

\begin{figure}[t!]
\includegraphics[width=1\columnwidth]{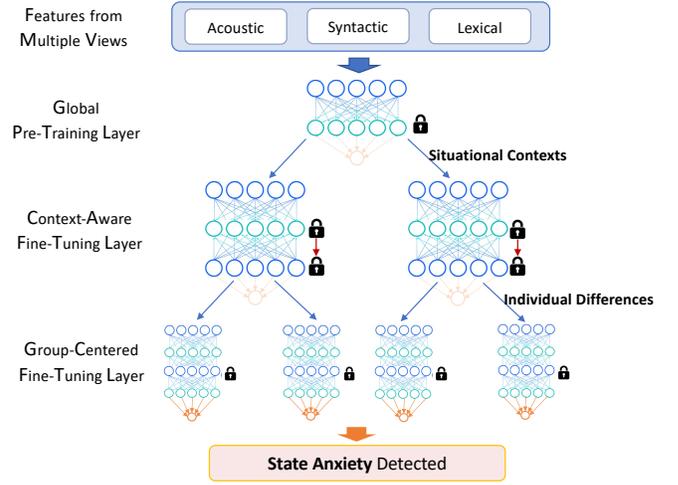}
\caption{Proposed multi-layer personalized ML pipeline.}
\label{fig:model}
\end{figure}

As shown in Figure \ref{fig:model}, \ours includes two components:

\subsubsection{\textbf{Multiview Featurization and Fusion}}\label{sec:multiview} The pipeline first extracts multi-view biomarkers $FS$ from the raw input data stream $D$ while using domain knowledge. Then it fuses $FS$ from different aspects/views with different domain-specific meanings (e.g, in our case, acoustic, syntactic, and lexical features) into one hidden vector $H$. Specifically, to model the contributions of different views' inputs, we assign learnable coefficients $\alpha_i$ to each view $i$ and refine the coefficient weights during the training process; formally, 
\begin{equation}
    H = \sum_{i \in N} \alpha_{i} \cdot g_{i} (FS_{i}, \theta_{gi})
\end{equation}
where H denotes the fused vector of biomarkers, N is the number of different views, $g$ considered a set of neural networks that maps $FS$ from $\mathbb{R}^{m_i}$ to $\mathbb{R}^K$, $K$ is the output dimension of $H$, and the $\theta_{gi}$ are the model parameters.

\subsubsection{\textbf{Multilayer Personalized Training}} consists of a pre-trained global module and a local grouping-based fine-tuned module, leveraging of both population-based and personalized (i.e., accounting for situational contexts and individual differences) information from the data collection. Inspired by the idea of transfer learning \cite{bengio2012deep}, the training process incorporates one \emph{global pre-training} set of layers for learning from the population domain and two \emph{fine-tuning} layers for situational context and individual cohort domains, respectively. Firstly, in the global pre-training step. We train a global model $M_G$ (i.e., $M_G(\theta_{G}, H)$), where $\theta_{G}$ is the parameter set of the model and $H$ is all the samples, with output $\hat{S}$ based on the ground-truth state anxiety status $S$ using a binary cross entropy objective loss function $BCE(\hat{S}, S)$. After optimizing $BCE(\hat{S}, S)$, the model parameter set $\theta_G$ is frozen as a globally \emph{pre-trained} model for further training. Then, in the fine-tuning step, based on pre-trained $M_G$ with frozen parameter set $\theta_G$, we subsequently attach new layers to $M_G$ (without its output layer) to adapt the model to fine-tuned $M_L$ according to the specific clustered samples $H_p$ by situational contexts and then by individual differences, where $p \in P$, $P$ denotes the set of participants in a specific group. The resulting prediction is $\hat{S_p} = M_L(\theta_{L}, H_p)$.

\subsection{Model Evaluation} \label{section6-evaluation}
We evaluated \ours with the following research questions.  
\begin{itemize}
    \item \textbf{RQ1}: Does \ours outperform generic (non-personalized) ML models? 
    \item \textbf{RQ2}: How does the "multiview featurization and fusion" component optimize the model performance?
    \item \textbf{RQ3}: How does the "multilayer personalized training" component optimize the model performance?
\end{itemize}

We used the data collected in Study 1 to evaluate the ML pipeline. For the \emph{context-aware layer}, the observations were divided by the situational contexts (i.e., non-evaluative and evaluative); for the \emph{group-centered layer}, the sample was divided by high/low symptom severity subgroups. 

We compared our \ours with 5 ML methods, including K Nearest Neighbour (KNN), Support Vector Machine (SVM), Extreme Gradient Boosting (XGBoost), Multilayer Perceptron (MLP), Random Forest (RF) Classifier \cite{shailaja2018machine}. Evaluation metrics include accuracy, precision, and F1-Score.

We performed a cross-validation grid search to determine the best hyper-parameters (e.g., learning rate, and training epoch) for every baseline (i.e., each comparison method) and \ours in a leave-one-sample-out cross-validation (LOOCV) manner. To make it a fair evaluation, 1) for the model with neural networks (i.e., \ours and MLP), we set the number of layers to 4 (in \ours, 2 for global training, 1 for context-aware fine-tuning, 1 for group-centered fine-tuning; in MLP, 4 layers in total) to guarantee that they have the same learning capability; 2) situational (non-evaluative or evaluative) and individual (high or low symptom severity) information was embedded into the feature space of the baseline models.

\begin{table}[t]
\centering
\small
 \caption{\label{tab:maintable} Comparison of model performance between proposed \ours and the baselines (i.e., comparison methods).}
 \scalebox{0.85}
{
 \begin{tabular}{lccc}
 \hline
 &\multicolumn{3}{c}{Metrics (\%)}\\
 \cline{2-4}

\multirow{-2}{*}{Methods} & Accuracy & Precision 
& F1 Score \\
\hline
KNN & 50.90 & 50.94 
& 50.91 \\
SVM & 41.81 & 41.79 
& 41.81 \\
XGBoost & 58.18 & 58.36 & 58.09 \\
MLP & 58.18 & 58.22 
& 58.18 \\
RF & 56.76 & 56.86 
& 56.72 \\
\hline
\ours & \textbf{74.55} & \textbf{74.64} 
& \textbf{74.49}\\
\hline
\end{tabular}}
\end{table}

\subsubsection{RQ1}

As shown in table \ref{tab:maintable}, \ours shows improvements on every metric. Specifically, \ours scored 74.55\% in accuracy, 74.64\% in precision, and 74.49\% in F1-score, which improved at least 28.0\% compared to baseline models. This indicates that the \ours method can better detect the state anxiety status of highly socially anxious people.

We also compared \ours to separately trained models for 4 subgroups to further evaluate its effectiveness. We found that \ours performed significantly better in evaluative contexts and had a higher overall performance than the separate models. As shown in Table \ref{tab:seperate}, our multilayer design showed a 10.85\% improvement in performance due to its use of a hierarchical multilayer design.

\begin{table}[t]
\small
 \caption{\label{tab:seperate} \ours vs separately trained models}
\scalebox{0.85}
{
 \begin{tabular}{lcccccc}
 \hline
 &\multicolumn{3}{c}{\ours}& \multicolumn{3}{c}{Separated}\\
\cline{2-4} \cline{5-7}
\multirow{-2}{*}{Subgroups} & Acc & Prec & F1 & Acc & Prec & F1\\
\hline
High, Non-Eval & 71.43 &  71.43  &  71.43 & 100.00&  100.00&    100.00\\
Low, Non-Eval  & 52.91 & 52.38 & 52.16 & 61.90 &   61.90  &  61.90 \\
High, Eval & 100.00 &  100.00  & 100.00 & 61.73 & 61.11  & 61.23\\
Low, Eval & 69.10 & 69.09 & 69.07 & 45.56 & 44.44 & 44.44\\
\hline
Overall & \textbf{74.55} & \textbf{74.64} & \textbf{74.49} & 63.64 &   63.64 & 63.64\\
\hline
\end{tabular}}

\end{table}

\subsubsection{RQ2}

\begin{table}[t]
\centering
\caption{\ours without (w/o) every single view.}
\scalebox{0.85}
\small
 \label{tab:dataview}
{
 \begin{tabular}{lccc}
 \hline
 &\multicolumn{3}{c}{Metrics (\%)}\\
\cline{2-4}
\multirow{-2}{*}{Views} & Accuracy & Precision 
& F1 Score \\
\hline
w/o Acoustic & 62.35 & 61.82 & 61.56\\
w/o Syntactic & 52.80 & 52.73 & 52.70\\
w/o Lexical & 49.15  & 49.09 & 49.06\\
\hline
\ours & \textbf{74.55} & \textbf{74.64} & \textbf{74.49}\\
\hline
\end{tabular}}
\end{table}

We conducted a feature ablation study to understand the effect of each linguistic view on model performance. We removed each view (acoustic, syntactic, and lexical) from the feature space and re-trained the model. The results, shown in Table \ref{tab:dataview}, indicate that the model performed best when equipped with all the features. Additionally, the learnable coefficient $\alpha$ for each modality shows the importance of each view of data: $\alpha_{acoustic}$ = 0.5071, $\alpha_{syntactic}$ = 0.5801, $\alpha_{lexical}$ = 0.8772. Both the ablation study and the coefficient $\alpha$ suggest that the lexical perspective is the most indicative of state anxiety, followed by the syntactic.

\subsubsection{RQ3}

\begin{table}[t]
\centering
\small
 \caption{\ours without (w/o) designed layers.}
 
 \label{tab:multilevel}
\scalebox{0.85}
{
 \begin{tabular}{lccc}
 \hline
 &\multicolumn{3}{c}{Metrics (\%)}\\
\cline{2-4}
\multirow{-2}{*}{Methods} & Accuracy & Precision 
& F1 Score \\
\hline
Only Global Layers & 58.18 & 58.22 & 58.18\\
w/o Context-Aware Layers & 63.66 & 63.64 & 63.56\\
w/o Group-Centered Layer & 68.01 & 67.27 & 66.81\\
\hline
All Layers (\ours) & \textbf{74.55} & \textbf{74.64} & \textbf{74.49}\\
\hline
\end{tabular}}
\end{table}

To investigate the effect of the multilayer fine-tuning step, we conducted a second ablation study to understand how the context-aware and group-centered layers contribute to model performance. We removed one or two sublayers at a time and re-trained the model. The results, shown in Table \ref{tab:multilevel}, indicate that both the context-aware and group-centered considerations contribute to the final detection performance. However, the context-aware layer is more critical than the group-centered layer as the F1 score dropped to 63.56\% when it was removed, compared to 66.81\% when the group-centered layer was removed.

\section{Discussion} \label{section7-discussion}

\subsection{Summary of Findings} \label{discussion-summary}

Study 1 indicated that linguistic patterns vary across social contexts and individual symptom profiles. Specifically, we observed that participants had lower pitch and longer sentences on average in evaluative contexts, which is aligned with prior human behavior research \cite{leongomez2017perceived}. Moreover, we found that individuals with higher anxiety symptom severity tended to exhibit specific linguistic characteristics, such as lower energy, zero-crossing rate, and pitch, longer sentences, and more stop words, particularly in an evaluative context. These patterns suggest that some groups' specific behavioral patterns are more likely to be associated with some social contexts than others, pointing to a person by context interaction (e.g., individuals with higher symptom severity may struggle to regulate emotions in contexts involving greater social threat \cite{sonnenschein2018linguistic}).

Study 2 advanced the ML model to be more personalized and learn more efficiently. Specifically, with \ours, ML models can learn more personalized knowledge from limited human datasets in a hierarchical learning fashion, especially when highly socially anxious people are in an evaluative social threat context (RQ1). Both context-aware and group-centered information are crucial for the model's performance in digital state anxiety detection. In particular, the ablation study showed that the context-aware layer plays a more vital role in this detection (RQ2). Notably, acoustic, syntactic and lexical features all contribute to effective state anxiety detection. Results show that lexical features are the most indicative of state anxiety, which can inform future digital mental health practice (RQ3).

\subsection{Implications for Future Practice} \label{discussion-implications}

The present findings have multiple implications for state anxiety detection. First, linguistic characteristics are indicative of individuals' state anxiety. This, in turn, suggests that researchers working to build JITAIs may benefit from collecting linguistic information to identify opportune moments for intervention. Of course, determining how to do this in ways that are acceptable to the user and their contacts in terms of privacy, confidentiality, intrusiveness and other ethical issues is paramount. Second, personalizing prediction models by using information about the individual's emotional health and symptoms, and their current social context helps us better detect the phenomenon of interest. Thus, future JITAI work would likely benefit from incorporating both individual differences and contextual features into prediction models. Third, our findings indicate that there is heterogeneity in terms of how linguistic features relate to state anxiety. That is, not only did linguistic features vary across evaluative and non-evaluative contexts and across symptom severity groups, but accounting for this variability in personalized models aided our prediction of state anxiety. This is in line with prior work suggesting that emotional states may not have a distinct 'fingerprint' and may instead relate to physiological data differently based on features of the individual in a particular situation \cite{siegel2018emotion}. At the same time, our findings suggest that clustering individuals based on psychological constructs (achieving a middle-ground between nomothetic and idiographic approaches) allows us to obtain valuable information about state anxiety beyond individuals. Future work aimed at detecting potential JITAI targets would benefit from considering cluster-based semi-idiographic methods to optimize prediction.

\subsection{Limitations and Future Directions} \label{discussion-limitations}

Our study has several limitations, which point to opportunities for future research. Our analyses were limited to the two social contexts and the two subgroups among a small sample of university students. As such, our findings may not generalize to other individual differences, contextual features, and populations. Future work should examine these questions across different contexts (e.g., in-person vs. virtual interactions) and individual differences (e.g., gender identity, age). Also, as noted, even though linguistic data may help researchers improve state anxiety detection, there are many privacy concerns which could impact participants' willingness to use the technology. This is a concern researchers should keep in mind and they should ensure they are using secure methods, discussing concerns with participants, etc.

\section{Conclusion} \label{section8-conclusion}

This paper tested the ability of personalized (vs. one-size-fits-all) ML approaches to detect state anxiety from linguistic biomarkers. Then, we proposed a personalized ML pipeline which progressively trains the model according to different domain knowledge (i.e., contextual and subgroup). We believe our personalized method may have considerable clinical utility relative to nomothetic ML approaches, and provide novel insights into how to optimize  detection of key mental health outcomes.

\bibliographystyle{IEEEtrans}
\bibliography{reference}

\begin{thebibliography}{10}
\providecommand{\url}[1]{#1}
\csname url@samestyle\endcsname
\providecommand{\newblock}{\relax}
\providecommand{\bibinfo}[2]{#2}
\providecommand{\BIBentrySTDinterwordspacing}{\spaceskip=0pt\relax}
\providecommand{\BIBentryALTinterwordstretchfactor}{4}
\providecommand{\BIBentryALTinterwordspacing}{\spaceskip=\fontdimen2\font plus
\BIBentryALTinterwordstretchfactor\fontdimen3\font minus
  \fontdimen4\font\relax}
\providecommand{\BIBforeignlanguage}[2]{{%
\expandafter\ifx\csname l@#1\endcsname\relax
\typeout{** WARNING: IEEEtran.bst: No hyphenation pattern has been}%
\typeout{** loaded for the language `#1'. Using the pattern for}%
\typeout{** the default language instead.}%
\else
\language=\csname l@#1\endcsname
\fi
#2}}
\providecommand{\BIBdecl}{\relax}
\BIBdecl

\bibitem{kessler2012twelve}
R.~C. Kessler, M.~Petukhova, N.~A. Sampson, A.~M. Zaslavsky, and H.-U.
  Wittchen, ``Twelve-month and lifetime prevalence and lifetime morbid risk of
  anxiety and mood disorders in the united states,'' \emph{International
  journal of methods in psychiatric research}, vol.~21, no.~3, pp. 169--184,
  2012.

\bibitem{hofmann2007cognitive}
S.~G. Hofmann, ``Cognitive factors that maintain social anxiety disorder: A
  comprehensive model and its treatment implications,'' \emph{Cognitive
  behaviour therapy}, vol.~36, no.~4, pp. 193--209, 2007.

\bibitem{grant2005epidemiology}
B.~F. Grant, D.~S. Hasin, C.~Blanco, F.~S. Stinson, S.~P. Chou, R.~B.
  Goldstein, D.~A. Dawson, S.~Smith, T.~D. Saha, and B.~Huang, ``The
  epidemiology of social anxiety disorder in the united states: results from
  the national epidemiologic survey on alcohol and related conditions,''
  \emph{Journal of Clinical Psychiatry}, vol.~66, no.~11, pp. 1351--1361, 2005.

\bibitem{wang2005failure}
P.~S. Wang, P.~Berglund, M.~Olfson, H.~A. Pincus, K.~B. Wells, and R.~C.
  Kessler, ``Failure and delay in initial treatment contact after first onset
  of mental disorders in the national comorbidity survey replication,''
  \emph{Archives of general psychiatry}, vol.~62, no.~6, pp. 603--613, 2005.

\bibitem{nahum2018just}
I.~Nahum-Shani, S.~N. Smith, B.~J. Spring, L.~M. Collins, K.~Witkiewitz,
  A.~Tewari, and S.~A. Murphy, ``Just-in-time adaptive interventions (jitais)
  in mobile health: key components and design principles for ongoing health
  behavior support,'' \emph{Annals of Behavioral Medicine}, vol.~52, no.~6, pp.
  446--462, 2018.

\bibitem{orru2020machine}
G.~Orr{\`u}, M.~Monaro, C.~Conversano, A.~Gemignani, and G.~Sartori, ``Machine
  learning in psychometrics and psychological research,'' \emph{Frontiers in
  psychology}, vol.~10, p. 2970, 2020.

\bibitem{hofmann2012linguistic}
S.~G. Hofmann, P.~M. Moore, C.~Gutner, and J.~W. Weeks, ``Linguistic correlates
  of social anxiety disorder,'' \emph{Cognition \& emotion}, vol.~26, no.~4,
  pp. 720--726, 2012.

\bibitem{dadi2021population}
K.~Dadi, G.~Varoquaux, J.~Houenou, D.~Bzdok, B.~Thirion, and D.~Engemann,
  ``Population modeling with machine learning can enhance measures of mental
  health,'' \emph{GigaScience}, vol.~10, no.~10, 2021.

\bibitem{wang2022personalized}
Z.~Wang, H.~Xiong, J.~Zhang, S.~Yang, M.~Boukhechba, D.~Zhang, L.~E. Barnes,
  and D.~Dou, ``From personalized medicine to population health: A survey of
  mhealth sensing techniques,'' \emph{IEEE Internet of Things Journal}, 2022.

\bibitem{olfson2000barriers}
M.~Olfson, M.~Guardino, E.~Struening, F.~R. Schneier, F.~Hellman, and D.~F.
  Klein, ``Barriers to the treatment of social anxiety,'' \emph{American
  Journal of Psychiatry}, vol. 157, no.~4, pp. 521--527, 2000.

\bibitem{vabalas2019machine}
A.~Vabalas, E.~Gowen, E.~Poliakoff, and A.~J. Casson, ``Machine learning
  algorithm validation with a limited sample size,'' \emph{PloS one}, vol.~14,
  no.~11, p. e0224365, 2019.

\bibitem{lovibond1996manual}
S.~H. Lovibond and P.~F. Lovibond, \emph{Manual for the depression anxiety
  stress scales}.\hskip 1em plus 0.5em minus 0.4em\relax Psychology Foundation
  of Australia, 1996.

\bibitem{ruscio2008social}
A.~M. Ruscio, T.~A. Brown, W.~T. Chiu, J.~Sareen, M.~B. Stein, and R.~C.
  Kessler, ``Social fears and social phobia in the usa: results from the
  national comorbidity survey replication,'' \emph{Psychological medicine},
  vol.~38, no.~1, pp. 15--28, 2008.

\bibitem{mattick1998social}
R.~P. Mattick and J.~C. Clarke, ``Social interaction anxiety scale,''
  \emph{Psychological Assessment}, 1998.

\bibitem{leary1983brief}
M.~R. Leary, ``A brief version of the fear of negative evaluation scale,''
  \emph{Personality and social psychology bulletin}, vol.~9, no.~3, pp.
  371--375, 1983.

\bibitem{kaufman2016difficulties}
E.~A. Kaufman, M.~Xia, G.~Fosco, M.~Yaptangco, C.~R. Skidmore, and S.~E.
  Crowell, ``The difficulties in emotion regulation scale short form (ders-sf):
  Validation and replication in adolescent and adult samples,'' \emph{Journal
  of Psychopathology and Behavioral Assessment}, vol.~38, no.~3, pp. 443--455,
  2016.

\bibitem{sonnenschein2018linguistic}
A.~R. Sonnenschein, S.~G. Hofmann, T.~Ziegelmayer, and W.~Lutz, ``Linguistic
  analysis of patients with mood and anxiety disorders during cognitive
  behavioral therapy,'' \emph{Cognitive behaviour therapy}, vol.~47, no.~4, pp.
  315--327, 2018.

\bibitem{ramchoun2016multilayer}
H.~Ramchoun, Y.~Ghanou, M.~Ettaouil, and M.~A. Janati~Idrissi, ``Multilayer
  perceptron: Architecture optimization and training,'' 2016.

\bibitem{bengio2012deep}
Y.~Bengio, ``Deep learning of representations for unsupervised and transfer
  learning,'' in \emph{Proceedings of ICML workshop on unsupervised and
  transfer learning}.\hskip 1em plus 0.5em minus 0.4em\relax JMLR Workshop and
  Conference Proceedings, 2012, pp. 17--36.

\bibitem{shailaja2018machine}
K.~Shailaja, B.~Seetharamulu, and M.~Jabbar, ``Machine learning in healthcare:
  A review,'' in \emph{2018 Second international conference on electronics,
  communication and aerospace technology (ICECA)}.\hskip 1em plus 0.5em minus
  0.4em\relax IEEE, 2018, pp. 910--914.

\bibitem{leongomez2017perceived}
J.~D. Leong{\'o}mez, V.~R. Mileva, A.~C. Little, and S.~C. Roberts, ``Perceived
  differences in social status between speaker and listener affect the
  speaker's vocal characteristics,'' \emph{PloS one}, vol.~12, no.~6, p.
  e0179407, 2017.

\bibitem{siegel2018emotion}
E.~H. Siegel, M.~K. Sands, W.~Van~den Noortgate, P.~Condon, Y.~Chang, J.~Dy,
  K.~S. Quigley, and L.~F. Barrett, ``Emotion fingerprints or emotion
  populations? a meta-analytic investigation of autonomic features of emotion
  categories.'' \emph{Psychological bulletin}, vol. 144, no.~4, p. 343, 2018.

\end{thebibliography}

\end{document}